\documentclass[prd,aps,twocolumn,showkeys,nofootinbib]{revtex4-2}

\usepackage{amsmath}
\usepackage{amsfonts}
\usepackage{amssymb}	
\usepackage{graphicx}
\usepackage{bm}
\usepackage{color}
\usepackage{ulem}
\usepackage{commath}
\usepackage[T1]{fontenc}
\usepackage{lmodern}
\usepackage{xcolor}
\usepackage{hyperref}
\allowdisplaybreaks

\def\p{\partial}

\def\Msun{M_{\odot}}
\def\GMc2{G M_{\odot} c^{-2}}

\def\lm{{\ell m}}
\def\EOB{{\rm EOB}}

\def\lm{{\ell m}}

\def\de{\partial}
\def\lm{{\ell m}}

\def\r{{\hat{r}}}
\def\ph{\varphi}

\def\F{{\cal F}}

\newcommand\be{\begin{equation}}
\newcommand\ee{\end{equation}}

\def\Msun{M_\odot}
\def\Qo{Q_{\omega}}
\def\FISF{\mathcal{F}_\ph^{\rm 1SF}}
\def\FIISF{\mathcal{F}_\ph^{\rm 2SF}}
\def\FIIISF{\mathcal{F}_\ph^{\rm 3SF}}
\def\tG{\tilde{G}}

\def\TEOBResumS{\texttt{TEOBResumS}}

\def\TEOBResumSDali{\texttt{TEOBResumS-Dal\'i}}

\newcommand{\aDJS}{\rm \overline{DJS}}

\begin{document}

\title{Comparing second-order gravitational self-force and effective-one-body waveforms from inspiralling, quasi-circular black hole binaries with a non-spinning primary and a spinning secondary}
\author{Angelica \surname{Albertini}${}^{1,2}$}
\author{Alessandro \surname{Nagar}${}^{3,4}$}
\author{Josh \surname{Mathews}${}^{5}$}
\author{Georgios \surname{Lukes-Gerakopoulos}${}^1$}

\affiliation{${}^1$Astronomical Institute of the Czech Academy of Sciences,
Bo\v{c}n\'{i} II 1401/1a, CZ-141 00 Prague, Czech Republic}
\affiliation{${}^2$Faculty of Mathematics and Physics, Charles University in Prague, 18000 Prague, Czech Republic}
\affiliation{${}^3$INFN Sezione di Torino, Via P. Giuria 1, 10125 Torino, Italy} 
\affiliation{${}^4$Institut des Hautes Etudes Scientifiques, 91440 Bures-sur-Yvette, France}
\affiliation{${}^5$Department of Physics, National University of Singapore, 21 Lower Kent Ridge Rd, Singapore 119077}

\begin{abstract}
We present the first comparison of waveforms evaluated using the effective-one-body (EOB) approach and gravitational self-force (GSF) theory for inspiralling black hole binaries with a non-spinning primary and a spinning secondary. This paper belongs to a series of papers comparing the EOB model \TEOBResumS{} to GSF results, where the latter are used to benchmark the EOB analytical choices in the large-mass-ratio regime. In this work, we explore the performance of two gauge choices for the gyro-gravitomagnetic functions $G_S$, $G_{S_*}$ entering the spin-orbit sector within the EOB dynamics. In particular, we consider the usual gauge of \TEOBResumS{}, where $G_S$ and $G_{S_*}$ only depend on the inverse radius and the radial momentum, and a different gauge where these functions also depend on the azimuthal momentum. The latter choice allows us to exploit as prefactor in $G_{S_*}$ the complete expression $G_{S_*}^K$ for a spinning particle on Kerr. As done previously, we employ both waveform alignments in the time domain and a gauge-invariant frequency-domain analysis to gain a more complete understanding of the impact of the new analytical choice. The frequency-domain analysis is particularly useful in confirming that the gyro-gravitomagnetic functions in the new chosen gauge bring the EOB spin contribution at 1st post-adiabatic order closer to the GSF one.
We finally implement the improved functions within the public code for \TEOBResumSDali{}, which already incorporates eccentricity. In this way, we upgrade the EOB model for extreme-mass-ratio inspirals presented in our previous work.
\end{abstract}

\date{\today}

\maketitle

\section{Introduction}
In roughly a decade from now we will witness the launch of a new generation of gravitational-wave (GW) detectors: either ground-based, like Einstein Telescope \cite{ET} and Cosmic Explorer \cite{Reitze:2019iox},  or orbiting in space, like the Laser Interferometer Space Antenna \cite{Audley:2017drz} and the DECi-hertz Interferometer Gravitational wave Observatory \cite{Kawamura:2020pcg}.  These planned interferometers will have improved sensitivities and will span over parts of the GW spectrum that are yet unobserved, allowing us to receive  signals from a large variety of sources, including new ones. Among the new sources, there are two we are interested in: intermediate- and extreme-mass-ratio inspirals (IMRIs and EMRIs respectively). These systems are composed of a primary massive or super-massive black hole and of a secondary less massive compact object, where the mass ratio of the primary with the respect to the secondary ranges from $\sim10^2$ to $\sim10^7$, hence they are generically called large-mass-ratio inspirals. The gravitational waveform emitted during the inspiral of the secondary object towards the primary tracks the spacetime around the primary black hole, thus constituting an excellent laboratory to test deviations from general relativity \cite{Babak:2017tow}.

To perform such tests, but also to detect the signal and infer the parameters of the source in the first place, fast and accurate theoretical templates are needed. As for detection, the key feature is the speed in evaluating the waveform, and with this target several models have already been developed, like the \texttt{FastEMRIWaveforms} package~\cite{Chua:2020stf, Katz:2021yft}, augmented kludge waveforms~\cite{Chua:2017ujo}, the tool \texttt{bhpwave}~\cite{Nasipak:2023kuf} for adiabatic inspirals into rotating black holes, and the recent work~\cite{Loutrel:2024qxp}, which analytically solves the post-Newtonian (PN) spin precession equations of motion in the EMRI limit.
To gain in speed these models have to give up some of the physics that comes into play in IMRIs and EMRIs, with a consequential loss in accuracy. Therefore, they might not be able to precisely infer all the parameters. 

As for physical completeness, one of the most flexible frameworks to work with is provided by the effective-one-body (EOB) approach~\cite{Buonanno:1998gg,Buonanno:2000ef,Damour:2000we,Damour:2001tu,Damour:2015isa}, that at given PN orders maps the two-body problem into the motion of a single body in an effective metric. This map provides the Hamiltonian, that accounts for the conservative sector and is complemented by an analytical radiation reaction. This formalism exploits PN expressions, frequently resumming them to improve their behaviour in the strong-field regime, but also needs to be tuned and benchmarked to exact results. As for comparable-mass binaries, the free parameters in EOB models are calibrated to numerical relativity (NR)~\cite{Nagar:2019wds, Nagar:2020pcj}, and the reliability of the waveforms is finally assessed in terms of the mismatch with respect to NR simulation catalogs~\cite{Riemenschneider:2021ppj, Albertini:2021tbt, Nagar:2022icd, Nagar:2023zxh}. 

When it comes to largely asymmetric binaries, NR can provide really few results\footnote{See however Refs.~\cite{Lousto:2022hoq,Nagar:2022icd,Rosato:2021jsq,Lousto:2020tnb} for 
ground-breaking studies that push NR simulations up to mass ratios 1000:1.}, and it is more natural to benchmark EOB models with waveforms evaluated via gravitational-self-force (GSF) theory. In a series of previous works~\cite{Nagar:2022fep, Albertini:2022dmc, Albertini:2023xcn}, we adapted the EOB model \TEOBResumS{} so that it could efficiently describe the waveform emitted by large-mass-ratio inspirals. In particular, both the conservative and the dissipative sectors were modified in order to match the second-order GSF (2GSF) results of Ref.~\cite{Wardell:2021fyy}  for nonspinning quasi-circular binaries. In Ref.~\cite{Albertini:2023xcn}, we presented the implementation of the improved functions into the publicly available code for \TEOBResumSDali{}, which includes aligned spins and eccentricity.

As another step in our ongoing effort to improve \TEOBResumS{} for large mass ratios, we look into the contribution of the secondary spin, since the availability of 2GSF results for black hole binaries with a non-spinning primary and spinning secondary~\cite{Mathews:2021rod} makes it possible for us to benchmark our choices. As for spin-related functions, we are aware the EOB model might need changes both in the dissipative and in the conservative sector.
Targeting the dissipative spin contribution is a delicate matter, and would require a study that starts from the insightful results of~\cite{Nagar:2019wrt} and updates it, possibly making use of the spinning secondary 2GSF fluxes. Such a task will be taken care of in future work. 
On the other hand, improving the spin contribution to the conservative sector is an easier task and is precisely the aim of the present work. As done in Refs.~\cite{Albertini:2022dmc, Albertini:2023xcn} we carry on analyses both in the time and in the frequency domain, and eventually rely on the second one to gain a clearer understanding.

In particular, we focus on the gyro-gravitomagnetic functions entering the spin-orbit sector of the EOB Hamiltonian, and compare two different spin gauges: the standard one~\cite{Rettegno:2019tzh} and a new one~\cite{Placidi:2024yld}.
While the outcome of the time-domain waveform alignment does not indicate which gauge yields a better agreement with GSF data, the gauge-invariant frequency-domain analysis provides a clearer answer, favoring the new one. 
Hence, we implement the new expression of the gyro-gravitomagnetic functions into the public eccentric branch of our \TEOBResumSDali{} model for large-mass-ratio inspirals.

The rest of this paper is organized as follows. Sec.~\ref{sec:GSF} briefs the GSF model employed to benchmark our results. The EOB formalism and the structure of \TEOBResumS{} are presented in Sec.~\ref{sec:EOB}, and Sec.~\ref{sec:Qomg} introduces our frequency-domain gauge-invariant analysis. In Sec.~\ref{sec:GSs}, we test the performance of the new gyro-gravitomagnetic functions by both time-domain and frequency-domain analyses, and we discuss our conclusions in Sec.~\ref{sec:conclusions}.

\textit{Notation and conventions:} We define the mass ratio as $q \equiv m_1/m_2 \ge 1$, the small mass ratio as $\epsilon \equiv m_2/m_1 \le 1$, the symmetric mass ratio $\nu\equiv m_1 m_2/M^2$, the ratios $X_{1,2} = m_{1,2}/M$, where $m_{1,2}$ are the masses of the two bodies and $M\equiv m_1+m_2$, while we use the convention $m_1\geq m_2$. The dimensionless spin variables are denoted as $\chi_{1,2}\equiv S_{1,2}/(m_{1,2})^2$, where $(S_1,S_2)$ are the individual, dimensionful, spin components along the direction of the orbital angular momentum. We use the geometric units with $G=c=1$ and typically normalize quantities by the total mass $M$, e.g. the time is $t\equiv T/M$, the radial separation $r\equiv R/M$ etc. Following the nomenclature in the self-force literature, we refer to quantities that make order-$\nu^{n-1}$ contributions to the orbital phase as ``$n$th post-adiabatic order'' ($n$PA).

\section{GSF results for binaries with a spinning secondary}
\label{sec:GSF}

Ref.~\cite{Mathews:2021rod} described how the effects of the secondary's spin may be added modularly to the 2GSF waveforms of Ref.~\cite{Wardell:2021fyy} for non-spinning binaries in a quasi-circular inspiral\footnote{Though the waveforms computed in Ref.~\cite{Mathews:2021rod} do not explicitly include the 2SF effects from Ref.~\cite{Wardell:2021fyy}.}. We combine these results to compute 2GSF waveforms including a spinning secondary. The resulting waveforms feature in the study in Ref.~\cite{Burke:2023lno}.

GSF waveform models rely upon the two-timescale expansion in which the small mass ratio $\epsilon$ of asymmetric mass binary systems (typically EMRIs and IMRIs) is exploited to separate the fast timescale of the orbital motion of the binary from the slow timescale, over which the orbital parameters evolve due to radiation reaction. In particular, the gravitational waveform model in this work follows from the two-timescale expansion of the Einstein field equations and the inspiral evolution scheme of Ref.~\cite{Miller:2020bft} which was extended to include the spin of the secondary in Ref.~\cite{Mathews:2021rod}. While the latter paper restricted the secondary's spin to be (anti-)aligned with the binary's orbital angular momentum, a following paper is in preparation allowing for a generic configuration of the secondary's spin and a slowly spinning primary \cite{Mathews:forwardcite}. The present work is still restricted to the (anti-)aligned spin configuration.
 
In Ref.~\cite{Albertini:2022rfe}, the differences between exact 1PA waveforms and several approximate models (namely the 1PAT1, 1PAT2 and 1PAF1 models from Ref.~\cite{Wardell:2021fyy}) were highlighted and ultimately comparisons between \TEOBResumS{} and 2GSF waveforms focussed on the 1PAT1 model. In our analysis we will also use the 1PAT1 approximation for the GSF waveforms which relies upon the following simplifications:
\begin{enumerate}
\item The evolution of the binary's orbital frequency is determined from a flux balance law with an approximated expression for the 2SF binding energy correction (see e.g. Ref.~\cite{Pound:2019lzj}), instead of being determined directly from the local self-force. In addition, the 2SF energy flux through the horizon is neglected in the balance law since it has not yet been computed, but it is expected to be small enough not to impact our findings~\cite{Martel:2003jj}.
\item The evolution of the primary's mass and spin is neglected. Despite formally contributing to the evolution of the waveform's phase at 1PA order, the impact of the evolution of the primary's mass and spin on the waveform are known to be numerically small~\cite{Wardell:2021fyy}.
\end{enumerate}
For more details regarding the omitted terms, refer to Section II A of Ref.~\cite{Albertini:2022rfe}. In the following subsections, we briefly describe the adaption of the 1PAT1 waveform model to include the secondary's spin.\\

\subsection{Orbital Dynamics}
Since we neglect the evolution of the primary's mass and spin, the key dynamical variable in the quasi-circular 1PAT1 model is the evolution of the orbital frequency which evolves according to;
\be
\label{eq:GSFOrbFreq}
\dfrac{d \Omega}{d s}=F^\Omega(\Omega(s)),
\ee
from which one determines the orbital phase via directly integrating the orbital frequency;
\be
\label{eq:GSFOrbPhase}
\phi_p(s)=\int^s \Omega(s') ds'.
\ee
In principle, when adding the secondary spin, one might expect an additional evolution equation for $\chi_2$. We highlight that $\chi_2$ is conserved at 1PA order and in the \TEOBResumS{} model. The two-timescale analysis for quasi-circular inspirals results in the determination of $F^\Omega$ as a series expansion in the small mass-ratio $\epsilon$:
\be
F^\Omega(\Omega)=\epsilon F^\Omega_{0}(\Omega)+ \epsilon^2 F_{1}^\Omega(\Omega)+O(\epsilon^3).
\ee
The secondary spin contributes linearly to the 1PA term and it is, thus, straightforward to include it in the evolution equation as
\be
 \epsilon^2 F_{1}^\Omega(\Omega)= \epsilon^2  F_{1,ns}^\Omega(\Omega)  + \epsilon^2 \chi_2  F_{1,s}^\Omega(\Omega).
 \ee
The functions $F^\Omega_{0}(\Omega)$, $F_{1,ns}^\Omega(\Omega)$  and $ F_{1,s}^\Omega(\Omega)$ are known and have been pre-computed via a flux balance formula (see e.g. Eq.~(3) of Ref.~\cite{Wardell:2021fyy} and Eq.~(74) of Ref.~\cite{Mathews:2021rod} respectively). The 0PA and the 1PA spin terms are exact, computed from the flux balance law from Ref.~\cite{Akcay:2019bvk} while the additional terms in the flux balance formula used to obtain $F_{1,ns}^\Omega(\Omega)$ in the 1PAT1 model are approximated \cite{Wardell:2021fyy, Albertini:2022rfe}.\\

\subsection{Waveform}
We decompose the waveform strain in the usual basis
\be
\label{eq:waveform}
h_+ - i h_\times = \dfrac{1}{\cal{D}_{\rm L}}\sum_{\ell=2}^{\ell_{\rm max}} \sum_{m=-\ell}^{\ell}h_\lm {}_{-2}Y_\lm(\iota,\Phi) \ ,
\ee
where $\cal{D}_{\rm L}$ is the luminosity distance,  ${}_{-2}Y_\lm(\iota,\Phi)$ are the $s=-2$ spin-weighted spherical harmonics and $\iota$ and $\Phi$ are the polar and azimuthal angle of the line of sight relative to the orbital plane. In the GSF formalism the waveform modes are constructed from a complex perturbation amplitude and the orbital phase:
\be
\label{eq:GSFwaveform}
h_\lm(s)= R_\lm(\Omega(s)) e^{- i m \phi_p(s)}.
\ee
The complex mode amplitudes are likewise known functions pre-computed from the Einstein equations as an expansion in powers of the mass-ratio
\be
\label{eq:ampexpand}
R_\lm(\Omega)=\epsilon R_\lm^{(1)}(\Omega)+ \epsilon^2 R_\lm^{(2,ns)}(\Omega)+\epsilon^2 \chi_2 R_\lm^{(2,s)}(\Omega)+\mathcal{O}(\epsilon^3).
\ee
Again the secondary's spin contributes linearly to the sub-leading amplitude and it is, thus, straightforward to add it to the existing non-spinning model.

\subsection{Re-expansion and model summaries}
Thus far, our small mass-ratio expansions have been in powers of $\epsilon$ at fixed values of $m_1$, $\Omega$ and $\chi_2$ which is a natural formulation from black hole perturbation theory. To facilitate comparisons with \TEOBResumS{}, we perform two similar re-expansions yielding two (slightly) different waveform models. In the first model, which we label 1PAT1-$a$, we re-expand in powers of the symmetric mass-ratio, $\nu$, at fixed values of $M$, $x \equiv (M \Omega)^{2/3}$ and $\tilde a_2\equiv X_2 \chi_2$. In the second model, which we label 1PAT1-$\chi$, we perform the same re-expansion but at fixed values of $\chi_2$ instead of $\tilde a_2$. This distinction is important when  performing expansions in $\nu$ as the relation between $\tilde a_2$ and $\chi_2$ is $\nu$-dependent. 

Expressing the series in terms of $\nu$ restores the system's symmetry under the interchange of $m_1$ and $m_2$, which in turn improves the accuracy of the expansion for systems with less extreme mass-ratios \cite{Warburton:2021kwk, Wardell:2021fyy, Albertini:2022rfe}. Similarly, the dependence on $\chi_2$ of the dynamics of generic mass-ratio binaries from `body 2' in Post-Newtonian theory are weighted by the mass-ratio $X_2$ (see e.g. Ref.~\cite{Messina:2018ghh}). This indicates that the quantity $\tilde a_2$ is a better parameter to hold fixed in the expansion in $\nu$ (as it was done in the 2GSF flux comparisons in Ref.~\cite{Warburton:2021kwk}) so that this explicit dependence is recovered at each individual order in the $\nu$ expansion. 

What then is the purpose of defining two different models? In general, we suggest that the 1PAT1-$a$ model more accurately captures the spin dynamics of the secondary as the mass-ratio becomes less extreme. This statement will be addressed more concretely in Ref.~\cite{Mathews:forwardcite} and this work will focus on comparisons with \TEOBResumS{} in the small mass-ratio regime. Meanwhile the 1PAT1-$\chi$ model is slightly more useful for comparisons of gauge invariant quantities at the level of individual terms in the power series (in $\nu$). To make this latter statement more precise, we will be computing the adiabaticity parameter, $\Qo(\omega)$, which is defined in Section~\ref{sec:Qomg}, and comparing the terms in the power series in $\nu$ from the GSF model and a fit to \TEOBResumS{}. The power series takes the form
\be
\label{eq:Qo_nu1}
\Qo = \frac{\Qo^{(0)}}{\nu}(\omega) + \Qo^{(1)}(\omega) + \mathcal{O}(\nu) .
\ee
In the 1PAT1-$\chi$ model, the 1PA term is independent of $\nu$;
\be
\Qo^{(1)}(\omega) =\Qo^{(1,ns)}(\omega)+\chi_2 \Qo^{(1,s)}(\omega),
\ee
while in the 1PAT1-$a$ model, the 1PA term is dependent on $\nu$;
\be
\Qo^{(1)}(\omega) =\Qo^{(1,ns)}(\omega)+\frac{\tilde a_2}{\nu} \Qo^{(1,s)}(\omega),
\ee
with $\tilde a_2\sim\nu \chi_2$. Thus, the 1PAT1-$\chi$ is more practical for comparing the values of $\Qo^{(1)}$, since we need only to make the comparisons over the two-dimensional parameter space $(\omega, \chi_2)$ as opposed to the three-dimensional parameter space $(\omega, \tilde a_2, \nu)$ with degeneracy between $\nu$ and $\tilde a_2$.

\subsubsection{The 1PAT-$a$ model summary.}

In terms of $\nu$, $x$ and $\tilde{a}_2$, the final 1PAT1-$a$ inspiral is determined via integrating
\begin{align}
\label{eq:ph-a}
\frac{d \phi_p}{ds} &= \Omega ,\\
\frac{d \Omega}{ds} &= \frac{\nu}{M^2} \left[ F_0(x) + \nu F_{1,ns}(x) +  \tilde a_2 F_{1,s}(x) \right] \label{eq:Omegadot}.
\end{align}
The functions $F_0(x)$ and $F_{1,ns}(x)$ are made explicit in Eq.~(12) and Eq.~(13) of Ref.~\cite{Albertini:2022rfe} (though there $F_{1,ns}(x)$ is simply labelled $F_{1}(x)$). The notation change from $F^\Omega_n$ to $F_n$ is to distinguish between coefficients of $\epsilon$ and coefficients of $\nu$, although the leading order term is simply $F_0(x)=F^\Omega_0(\hat\Omega(x))$ with $\hat \Omega\equiv M \Omega$. Similarly, the form of the spin term is  $F_{1,s}(x)= F_{1,s}^\Omega(\hat \Omega(x))$ and explicitly
\begin{multline}
F_{1,s}^\Omega(\hat \Omega(x))=\frac{3 x^{1/2} (1 - 3x)^{3/2}}{1-6x} \mathcal{F}^{(2,s)}(x)\\
 -\frac{3 (5-12 x) (1-3 x)^{3/2} x^2}{(1-6 x)^2} \mathcal{F}^{(1)}(x),
\end{multline}
which follows from the re-expansion of Eq.~(74) from Ref.~\cite{Mathews:2021rod}. The functions $\mathcal{F}^{(1)}$ and $ \mathcal{F}^{(2,s)}$ are the $\epsilon^1$ and $\epsilon^2 \chi_2$ coefficients respectively of the total gravitational energy flux radiated through the horizon and to infinity (when the flux is expanded in $\epsilon$ at fixed $\chi_2$).

The re-expanded waveform modes are
\begin{multline}\label{1PAT1hlm}
h_{\ell m} = \left[\nu R^{(1)}_{\lm}+\nu \tilde{a}_2 R^{(2,s)}_{\ell m}\right. \\
\left.+\nu^2\left(R^{(2,ns)}_{\ell m}+R^{(1)}_{\ell m}-\hat\Omega \partial_{\hat\Omega}R^{(1)}_{\ell m}\right)\right]e^{-im\phi_p},
\end{multline}
where $R^{(n)}_{\ell m}=R^{(n)}_{\ell m}(\hat\Omega)$ are the amplitudes in Eq.~\eqref{eq:ampexpand}. In the re-expansion of the amplitudes above, we have accounted for changing the luminosity distance from units of $m_1$ to units of $M$ by multiplying Eq.~\eqref{eq:ampexpand} by the mass-ratio $X_1$ to arrive at Eq.~\eqref{1PAT1hlm}.

\subsubsection{The 1PAT-$\chi$ model summary.}

In terms of $\nu$, $x$ and $\chi_2$, the 1PAT1-$\chi$ inspiral is determined via integrating
\begin{align}
\label{eq:ph-chi}
\frac{d \phi_p}{ds} &= \Omega ,\\
\frac{d \Omega}{ds} &= \frac{\nu}{M^2} \left[ F_0(x) + \nu F_{1,ns}(x) + \nu \chi_2 F_{1,s}(x) \right] \label{eq:Omegadotchi}.
\end{align}
Conveniently, all of the functions $F_0(x)$, $F_{1,ns}(x)$ and $F_{1,s}(x)$ are the same as in the 1PAT1-$a$ model in Eq.~\eqref{eq:Omegadot}. At the practical level at 1PA order, the only mathematical difference is the interchange $\tilde a_2 \leftrightarrow \nu \chi_2$. It is in the spin terms at 2PA order where the difference between to two spin models becomes more pronounced. Likewise the re-expanded waveform modes are
\begin{multline}\label{1PAT1hlmchi}
h_{\ell m} = \left[\nu R^{(1)}_{\lm}+\nu^2 \chi_2 R^{(2,s)}_{\ell m}\right. \\
\left.+\nu^2\left(R^{(2,ns)}_{\ell m}+R^{(1)}_{\ell m}-\hat\Omega \partial_{\hat\Omega}R^{(1)}_{\ell m}\right)\right]e^{-im\phi_p},
\end{multline}
where again the functions $R^{(n)}_{\ell m}=R^{(n)}_{\ell m}(\hat\Omega)$ et cetera are identical to those in Eq.~\eqref{1PAT1hlm}.

\section{The EOB model \TEOBResumS{}}
\label{sec:EOB}
As already mentioned we work with \TEOBResumS{}~\cite{Nagar:2018zoe,Nagar:2020pcj,Riemenschneider:2021ppj}, an EOB model originally built for comparable-mass black hole binaries. This model has been recently modified to increase its agreement with GSF results at large mass-ratios, as described in Refs.~\cite{Albertini:2022rfe, Albertini:2022dmc, Albertini:2023xcn}.
To have a self-contained work we brief in this section the key ingredients of the EOB formalism and of the structure of the model. 

Within the EOB approach, the evolution of a binary black hole system is mapped into the motion of a body with mass $\mu\equiv m_1 m_2/M$ in an effective metric. This map yields the Hamiltonian
\be
\hat{H}_{\rm EOB} \equiv \frac{H_{\rm EOB}}{\mu} = \frac{1}{\nu} \sqrt{1 + 2\nu \left(\hat{H}_{\rm eff} - 1\right)} ,
\ee
where the effective Hamiltonian $\hat{H}_{\rm eff}$ splits in an orbital and a spin-orbit contribution as:
\begin{align}
\hat{H}_{\rm eff} &= \hat{H}_{\rm eff}^{\rm orb} + p_\varphi \left(G_S \hat{S} + G_{S_*} \hat{S}_*\right) ,
\end{align}
where the orbital part reads:
\begin{align}
\hat{H}_{\rm eff}^{\rm orb} &= \sqrt{p_{r_*}^2 + A \left( 1 + p_{\varphi}^2 u_c^2 + Q \right)} . \label{eq:Horbeff}
\end{align}
Within the orbital contribution $p_{r_*} = (A/B)^{1/2}p_r$ is the conjugated momentum to the tortoise-rescaled radial coordinate $r_* = \int dr \, (A/B)^{1/2} $, $p_\varphi$ is the angular momentum conjugated to the orbital phase $\varphi$ and $u_c = 1/r_c$ is the inverse of the centrifugal radius\footnote{We neglect here the next-to-leading order spin contribution that is instead used in the comparable-mass \TEOBResumS{}.}, that is defined as \cite{Damour:2014sva} 
\be
r_c^2 = r^2 + \tilde{a}_0^2 \left( 1 + \frac{2}{r}\right) , 
\ee
where $\tilde{a}_0 = \chi_1 X_1 + \chi_2 X_2$. 

In the spin-orbit contribution, $G_S$ and $G_{S_*}$ are the gyro-gravitomagnetic functions and the spin variables are defined as $\hat{S} = (S_1 + S_2)/M^2$ and $\hat{S}_* = \frac{1}{M^2} \left( \frac{m_2}{m_1} S_1 +  \frac{m_1}{m_2} S_2 \right)$. In this framework $A, D \equiv AB$ and $Q$ are the three EOB potentials defining the conservative sector. In the version of \TEOBResumS{} we are using here, these potentials are considered up to linear order in the symmetric mass ratio $\nu$, and the linear-in-$\nu$ term includes a fit to numerical self-force data~\cite{Nagar:2022fep,Akcay:2012ea,Akcay:2015pjz}. As mentioned in Refs.~\cite{Nagar:2022fep, Albertini:2022dmc}, these potentials are singular at the light ring, so that
the current model is limited to inspirals only (but notably including plunge effects).

The gyro-gravitomagnetic functions account for the spin-orbit interaction. Their current expressions read
\begin{align}
G_S       &= G_S^0 \hat{G}_S, \quad \quad G_S^0 = 2uu_c^2 ,\\
G_{S_*} &= G_{S_*}^0 \hat{G}_{S_*}, \quad G_{S_*}^0 = (3/2) u_c^3 .
\end{align}
After the leading-order contributions $G_S^0, G_{S_*}^0$ are factorized out, the residuals are inversely resummed:
\begin{align}
\hat{G}_S &= \left( 1 + c_{10}u_c + c_{20}u_c^2 + c_{30}u_c^3 \right. \nonumber \\
		&\quad \quad \;\,  + c_{02}p_{r^*}^2 + c_{12}u_c p_{r^*}^2  + c_{04}p_{r^*}^4   \nonumber \\
		&\quad \quad \;\, \left. + c_{22}u_c^2 p_{r^*}^2 + c_{14}u_c p_{r^*}^4 + c_{06}p_{r^*}^6 \right) ^{-1}, \\
\hat{G}_{S_*} &= \left(1 + c^*_{10}u_c + c^*_{20}u_c^2 + c^*_{30}u_c^3 + c^*_{40}u_c^4 \right.  \nonumber \\
		     &\quad \quad \;\,  + c^*_{02}p_{r^*}^2 + c^*_{12}u_c p_{r^*}^2  + c^*_{04}p_{r^*}^4  \nonumber \\
		     &\quad \quad \;\, \left. + c^*_{22}u_c^2 p_{r^*}^2 + c^*_{14}u_c p_{r^*}^4 + c^*_{06}p_{r^*}^6 \right) ^{-1}, 
 \end{align}
where the analytical $\nu$-dependence is known for all the coefficients apart for $c^*_{40}$. In the model for comparable masses, the coefficients $c_{22}$, $c^*_{22}$, $c_{14}$, $c^*_{14}$, $c_{06}$, $c^*_{06}$ are set to zero, while in the coefficients $c_{30}$, $c^*_{30}$ the analytically-known $\nu$-dependence is replaced with an effective one through a parameter tuned to NR. Here, we turn off the NR parameter and use the available analytical information both for the coefficients that are set to zero for comparable masses and for $c_{30}$, $c^*_{30}$. This corresponds to using the next-to-next-to-next-to-leading order (N3LO) results of Ref.~\cite{Antonelli:2020aeb}. The analytical results would be functions of $u$, but in the standard version of \TEOBResumS{} $u$ is replaced with $u_c$, following Ref.~\cite{Damour:2014sva}. Expressions for $\hat{G}_S, \hat{G}_{S_*}$ can be found in Eq.~(45) in Ref.~\cite{Placidi:2024yld}. 

We highlight that these functions are are expressed in the Damour-Jaranowski-Sch\"afer (DJS) gauge, which is set by requiring that there is no dependence on $p_\varphi$ in $G_S$ and $G_{S_*}$~\cite{Damour:2000we, Damour:2008qf}. In this gauge, the prefactor $G_{S_*}^0$ corresponds to just the leading-order term in the PN expansion of the corresponding test-mass expression, while the prefactor $G_S^0$ coincides exactly with the test-mass expression of $G_S$. 

We remind that the necessity of choosing a spin gauge in EOB is connected to the choice of a center of mass of an extended body in a curved spacetime. This is clearly seen in Ref.~\cite{Damour:2008qf}, where to obtain the EOB (real) Hamiltonian for spinning binaries (with NLO spin-orbit coupling), one starts from the related ADM Hamiltonian in the center-of-mass frame and performs two canonical transformations. The first one is needed to go to EOB coordinates, while the second one only affects NLO spin-orbit terms, and involves two arbitrary $\nu$-dependent dimensionless coefficients. These coefficients can be viewed as gauge parameters, connected to the arbitrariness of defining a local frame to measure the spin vectors. The real EOB Hamiltonian is then related to the effective one via the usual energy map. In Ref.~\cite{Damour:2008qf} the authors also point out that, since the Hamiltonian in consideration is an approximation to the exact one, the choice of the spin gauge will still (weakly) affect physical results. The reader should bear this in mind when going through the results of the present paper. 

One can however choose another gauge, so as to be able to factorize out as $G_{S_*}^0$ the EOB generalization of the full gyro-gravitomagnetic function for a spinning particle on a Kerr background, which reads~\cite{Bini:2015xua}
\begin{align}
\label{eq:GSs0sp}
G_{S_*}^K &= \frac{1}{(r_c^K)^2} \left\{ \frac{\sqrt{A^K}}{\sqrt{Q^K}} \left[ 1 - \frac{(r_c^K)'}{\sqrt{B^K}} \right ] + \right. \nonumber \\
		& \left. + \frac{r_c^K}{2+(1 + \sqrt{Q^K})}\frac{(A^K)'}{\sqrt{A^K B^K}} \right\} ,
\end{align}
where the functions labeled with $K$ are those related to the Kerr spacetime, namely
\begin{subequations}
\begin{align}
(r_c^K)^2 &= r^2 + a^2 \left( 1 + \frac{2}{r}\right) , \label{eq:rcK} \\
u_c^K &= 1/r_c^K , \\
A^K & = \frac{(1 + 2 u_c^K) (1 - 2 u_c^K)}{1 - 2u} , \\
B^K &= \frac{(u_c^K)^2}{u^2 A^K} , \\
Q^K &= 1 + p_\varphi^2 (u_c^K)^2 + \frac{p_r^2}{B^K},
\end{align}
\end{subequations}
where in Eq.~\eqref{eq:rcK}, $a$ is the dimensionless Kerr parameter. Following Ref.~\cite{Rettegno:2019tzh}, we refer to such a gauge as anti-DJS ($\rm \overline{DJS}$). 

In this work, we exploit recently obtained results in this gauge at N3LO~\cite{Placidi:2024yld}. As proposed in Ref.~\cite{Rettegno:2019tzh}, for $G_{S_*}^0$ we take  expression~\eqref{eq:GSs0sp} and substitute instead of $r_c^K$, $A^K$ and $B^K$ the respective EOB functions\footnote{The function $Q^K$ is \textit{not} related to the EOB orbital potential $Q$.}, i.e. the EOB centrifugal radius and the EOB potentials $A, B$. We remind that the latter are taken at linear order in $\nu$ for this version of \TEOBResumS{} and incorporate 1st order self-force information.  As for the centrifugal radius, we make a different choice with respect to the standard model and set the NLO correction to zero, so to only have the LO contribution. This means the $r_c$ expression is analogous to the Kerr one but replacing $a$ with $\tilde{a}_0$. 
The other prefactor, $G_S^0$, remains the same as the DJS prescription. The new residuals\footnote{The residual $\hat{G}_{S_*}$ we write here has been actually obtained substituting in $G_{S_*}^0$ the EOB potentials at 5PN, with all the $\nu$-depedence for every PN order. This is in fact the standard choice for the comparable-mass EOB model. The potentials of the version of \TEOBResumS{} we are using here instead are, as already mentioned, linear in $\nu$, at 8.5PN~\cite{Nagar:2022fep}, and multiplied by a fit on self-force data that affects terms from 7PN on. On the one side, the new gyro-gravitomagnetic functions are at 4.5PN, so that the computation is not influenced by differences at higher PN orders. On the other side, the fact that the 5PN potentials do incorporate higher-order-in-$\nu$ contributions implies that the obtained $\hat{G}_{S_*}$ is influenced in the non-linear-in-$\nu$ dependence. We choose to keep this choice anyway since 
at a later stage in the development of this version of \TEOBResumS{}, we might include higher-order-in-$\nu$ analytical information in the potentials.}, again inversely resummed, read
\begin{align}
\hat{G}_S &= \left( 1 + c_{10}u + c_{20}u^2 + c_{30}u^3 \right. \nonumber \\
		&\quad \quad \;\,  + ( c_{02} + c_{12}u  + c_{22} u^2 ) \, p^2   \nonumber \\
		&\quad \quad \;\,  + ( c_{04} + c_{14}u  ) \, p^4   \nonumber \\
		&\quad \quad \;\, \left. + \; c_{06}p^6 \right) ^{-1}, \\
\hat{G}_{S_*} &= \left(1 + c^*_{100}u + (c^*_{200} + c^*_{202} p_{r_*}^2 )u^2 + c^*_{300}u^3 \right. \nonumber \\
		&\quad \quad \;\,  + ( c^*_{020} + (c^*_{120} + c^*_{122} p_{r_*}^2 ) u  + c^*_{220} u^2 ) \, p^2   \nonumber \\
		&\quad \quad \;\,  + ( c^*_{040} + c^*_{140} u  ) \, p^4   \nonumber \\
		&\quad \quad \;\, \left. + \; c^*_{060} p^6 \right) ^{-1}, 
 \end{align}
where all the coefficients are analytically known and can be found in Eq.~(48) in Ref.~\cite{Placidi:2024yld}. Differently from their DJS representation, these residuals are expressed as functions of $p^2 = p_r^2 + p_\varphi^2 u^2$, since in this gauge it is allowed for the gyro-gravitomagnetic functions to depend on $p_\varphi$. Moreover, we choose to keep the simple dependence on $u$ rather than $u_c$. 
In the following we will analyse the effect of using the $\aDJS$ gauge with the aim of obtaining an increased agreement with GSF results.

The Hamiltonian equations read
\begin{subequations} \label{eq:EOBmotion}
\begin{align} 
\dot{\ph} & = \p_{p_\ph} \hat{H}_\EOB \equiv \Omega, \\
\dot{r} &= \left( \frac{A}{B} \right)^{1/2} \p_{p_{r_*}} \hat{H}_\EOB, \\
\dot{p}_\ph &= \hat{\F}_\ph , \\
\dot{p}_{r_*} &=  \left( \frac{A}{B} \right)^{1/2} \left( - \p_{r} \hat{H}_\EOB + \hat{\F}_r \right)
\end{align}
\end{subequations}
where $\Omega$ is the orbital frequency and $\hat{\F}_\varphi, \hat{\F}_\r$ are respectively the azimuthal and the radial contributions to the radiation reaction force, accounting for  the gravitational wave emission both to infinity and into the black holes' horizons.  In the version of \TEOBResumS{} we use here, both the asymptotic contribution and the horizon one have been enhanced~\cite{Albertini:2022dmc, Albertini:2023xcn} to improve the agreement of the 0PA contribution of \TEOBResumS{} with the 2GSF results of Ref.~\cite{Wardell:2021fyy}. The spin contribution to the flux is the same one that is used for comparable-mass binaries~\cite{Nagar:2020pcj}.

Solving the set~\eqref{eq:EOBmotion} of ODEs provides us with the time evolution of the dynamical variables, that enters the evaluation of the waveform in Eq.~\eqref{eq:waveform}.
In the following, we will also work with the Regge-Wheeler-Zerilli (RWZ) normalization convention and express the waveform as $\Psi_\lm\equiv h_\lm/\sqrt{(\ell+2)(\ell+1)\ell(\ell-1)}$. The RWZ normalized strain quadrupole waveform is then separated into amplitude and phase with the convention
\be
\label{eq:RWZnorm}
\Psi_{22} (t)= A_{22}(t) e^{-i \phi_{22}(t)}.
\ee
The instantaneous gravitational wave frequency (in units of $M$) is defined as $\omega_{22} \equiv \dot{\phi}_{22}$.

\begin{figure*}[htp!]
\includegraphics[width=0.85\textwidth]{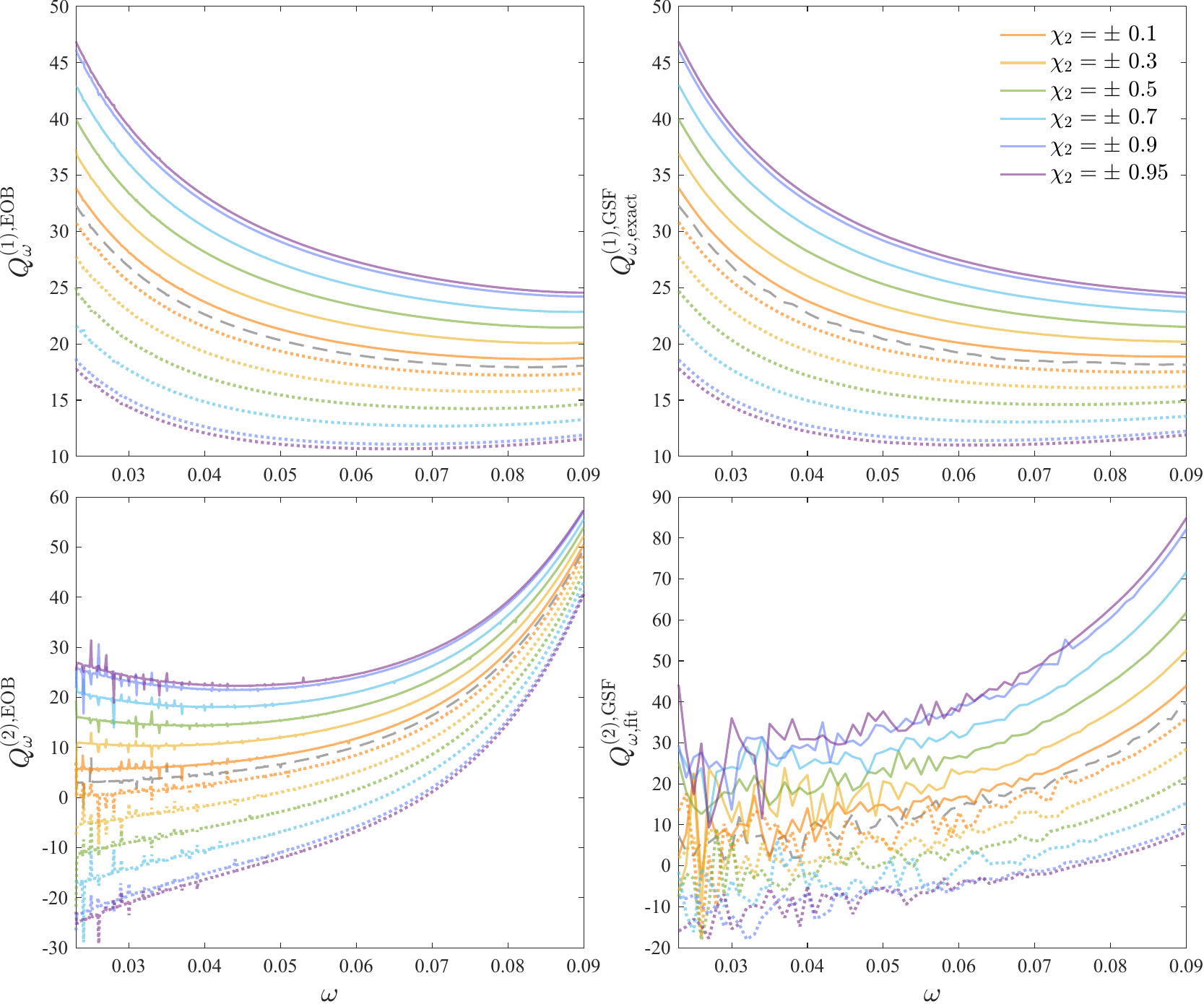}
\caption{\label{fig:Q12} The functions $\Qo^{(1)}$ and $\Qo^{(2)}$ from Eq.~\eqref{eq:Qo_nu} (top, bottom) for EOB and GSF (left, right). The nonspinning results are plotted in dashed grey, and dotted lines indicate negative values of the secondary spin. As for GSF, for $\Qo^{(1)}$ we use the exact function corresponding to the 1PAT1-$\chi$ model, while for $\Qo^{(2)}$ we use the result of the
fitting procedure described in the text. From these plots we can see that (i) $\Qo^{(1)}$ has a linear dependence on the secondary spin $\chi_2$, (ii) $\Qo^{(2)}$ has a non-linear dependence on $\chi_2$, that is stronger for GSF and (iii) as expected, positive spins induce a more adiabatic evolution (corresponding to a larger value of $\Qo$), while the opposite holds for negative spins.}
\end{figure*}

\begin{figure}
\includegraphics[width=0.48\textwidth]{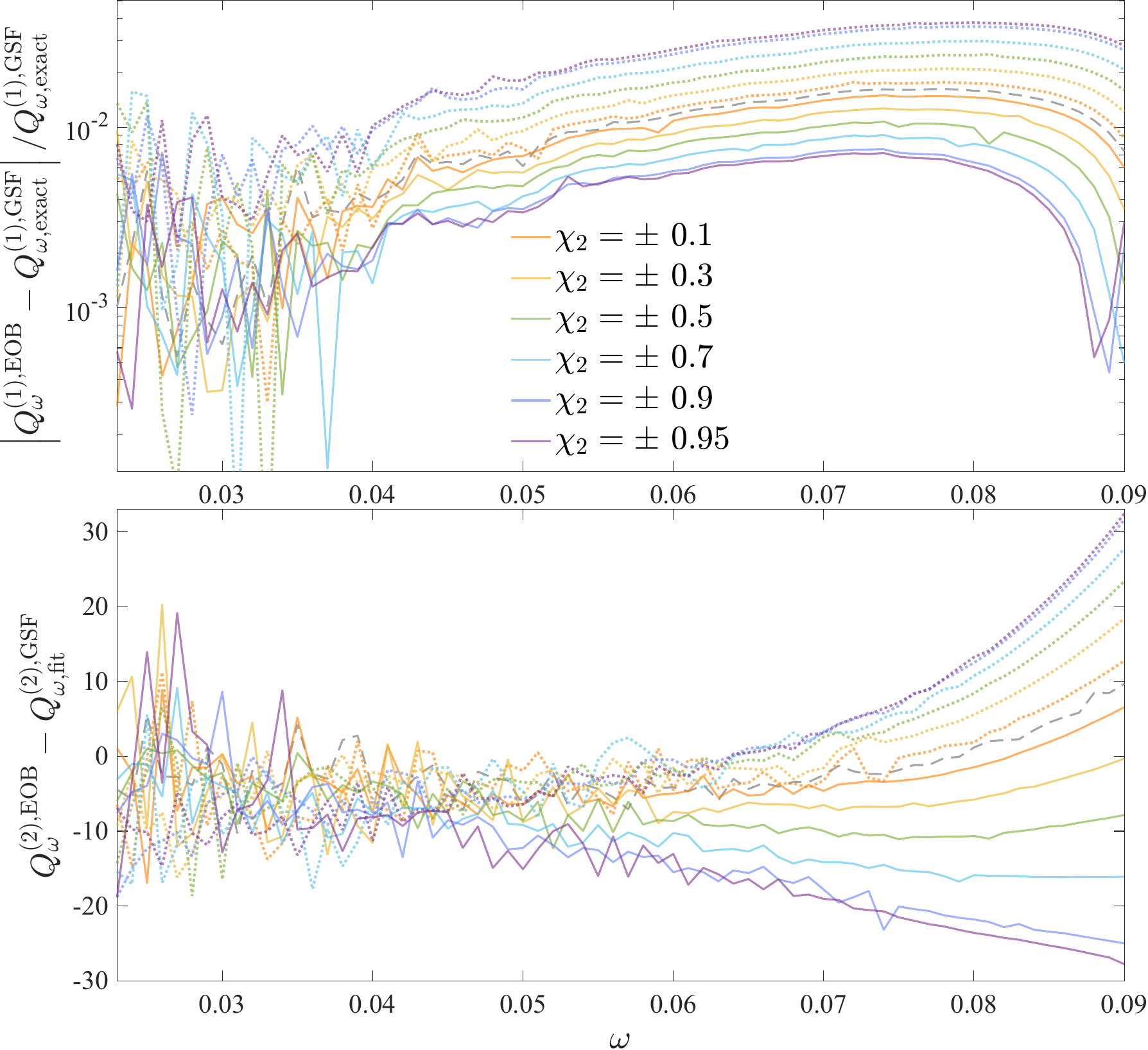} 
\caption{\label{fig:dQ12}  Relative difference in $\Qo^{(1)}$ (top) and absolute difference for $\Qo^{(2)}$ (bottom). We use the absolute difference for $\Qo^{(2)}$ since the function takes values near to zero for some values of $\chi_2$. Again, the GSF results displayed here correspond to the 1PAT1-$\chi$ model.}
\end{figure}
%

\section{A first gauge-invariant analysis}
\label{sec:Qomg}
Before exploring the impact of another gauge choice, in this section we test the performance of the EOB model as it is. Namely, we test it with the potentials and fluxes that have been previously improved for nonspinning binaries~\cite{Albertini:2022dmc, Albertini:2023xcn}, and use the spin-orbit functions in the DJS gauge described above, corresponding to the results of Ref.~\cite{Antonelli:2020aeb}.

 As in our previous work, we define the adiabaticity parameter
\be
\Qo = \frac{\omega^2}{\dot{\omega}},
\ee
where $\omega \equiv \omega_{22}$ is the $\ell = m = 2$ waveform frequency. Within the GSF approach, for a fixed value of $\omega$, $\Qo$ can be given as an expansion in $\nu$ (see Eq.~\eqref{eq:Qo_nu1}). As in our previous work, we consider the expansion up to 2PA:
\be
\label{eq:Qo_nu}
\Qo = \frac{\Qo^{(0)}}{\nu} + \Qo^{(1)} + \nu \Qo^{(2)} .
\ee
In Refs.~\cite{Albertini:2022dmc, Albertini:2023xcn} this parameter has been especially useful in highlighting how
the enhancement of the flux and the potentials brought the content of the EOB model at 0PA and at 1PA closer to the GSF one.

Spin enters the expansion of $\Qo$ at 1PA, so we neglect $\Qo^{(0)}$. We fit the remaining $\Qo^{(1,2)}$ terms using the same procedure as in Ref.~\cite{Albertini:2022rfe},
with mass ratios $q = \{ 26, 32, 64, 128, 500, 5000 \}$ at fixed values of $\chi_2 = \pm \{ 0.1, 0.3, 0.5, 0.7, 0.9, 0.95 \}$. Namely, we have a different fit for $\Qo^{(1,2)}(\omega)$ for every value of $\chi_2$. As done before~\cite{Albertini:2022rfe}, we choose the frequency interval $(0.023, 0.09)$ in order to be far enough from the point where the approximations underlying the GSF model break down. For $\Qo^{(1)}$, it is possible to extract from the GSF model the exact result, as done already in Ref.~\cite{Albertini:2022dmc}. In Fig.~\ref{fig:Q12}, we show the results of the fitting procedure for $\Qo^{(1,2), \rm EOB}$ and for $\Qo^{(2), \rm GSF}$, while for $\Qo^{(1), \rm GSF}$ we use the exact result. As for $\Qo^{(2), \rm GSF}$, there is no exact result, since from the theoretical point of view the model is truncated at 1PA order. In practice we truncate at 1PA when integrating Eqs.~\eqref{eq:ph-a} (or Eqs.~\eqref{eq:ph-chi}) to obtain the orbital phase. When constructing the waveform modes via Eq.~\eqref{eq:GSFwaveform}, we leave the combination of the complex amplitudes and the phase factor exact.Thus in computing $\Qo$ numerically from the waveform, one finds 2PA and higher terms (and there are spin squared terms in the numerical results for $\Qo^{(2), \rm GSF}$ et cetera). Such higher order terms are incomplete since they are missing key higher order GSF information and are somewhat meaningless without it. The top panel of Fig.~\ref{fig:dQ12} shows the relative EOB-GSF difference in $\Qo^{(1)}$, while the bottom panel shows the absolute difference for $\Qo^{(2)}$. 
We do not evaluate the relative difference for $\Qo^{(2)}$ since the function takes values
near to zero for some values of $\chi_2$, which would cause the relative error to diverge. Moreover, we again stress that the GSF result is incomplete and it should not be used as a reference. The aim of the next section is to see whether by changing the spin-orbit sector of the EOB model we can improve the agreement of its $\Qo^{(1)}$ term with the GSF one.

\begin{figure*}[htp!]
\includegraphics[width=0.45\textwidth]{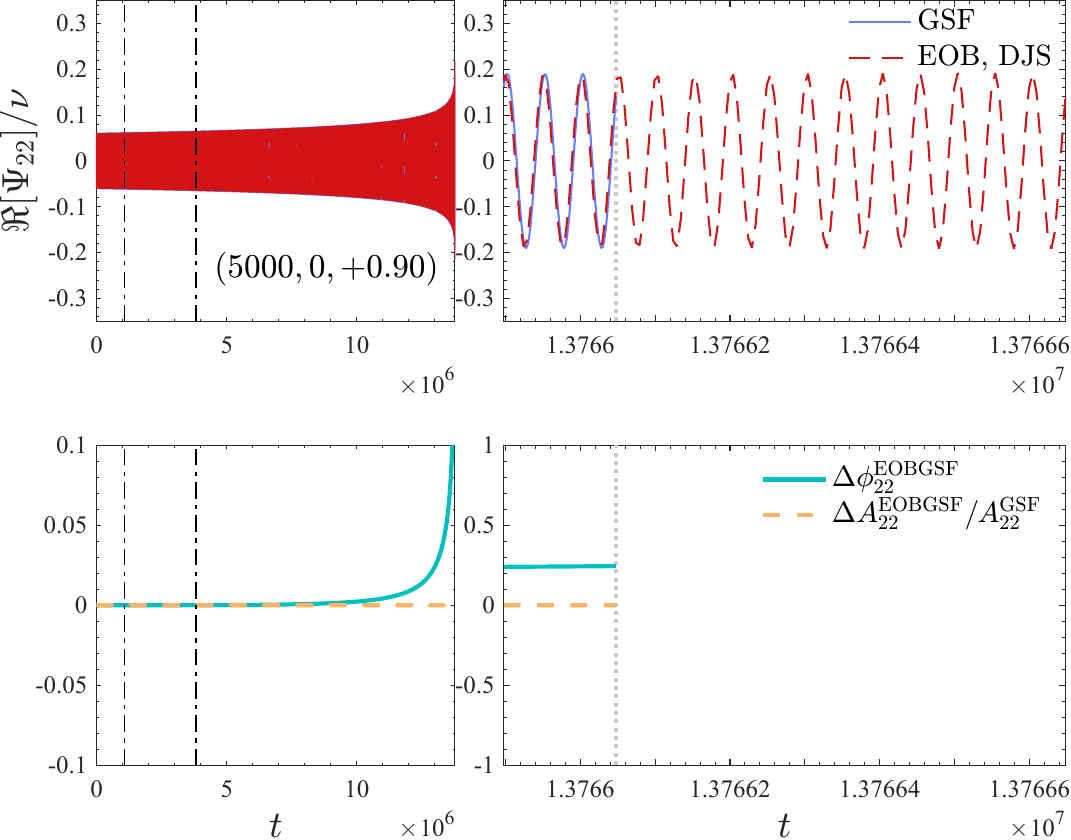}
\includegraphics[width=0.45\textwidth]{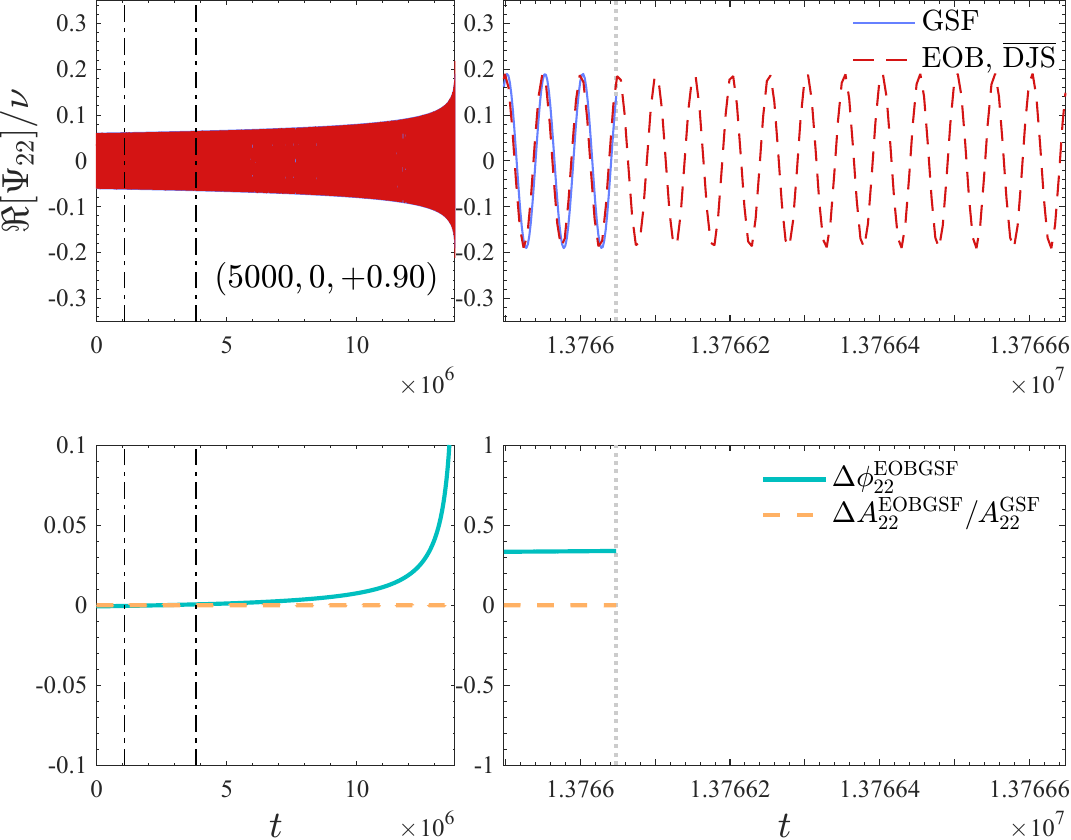} \\
\vspace{0.5mm}
\includegraphics[width=0.45\textwidth]{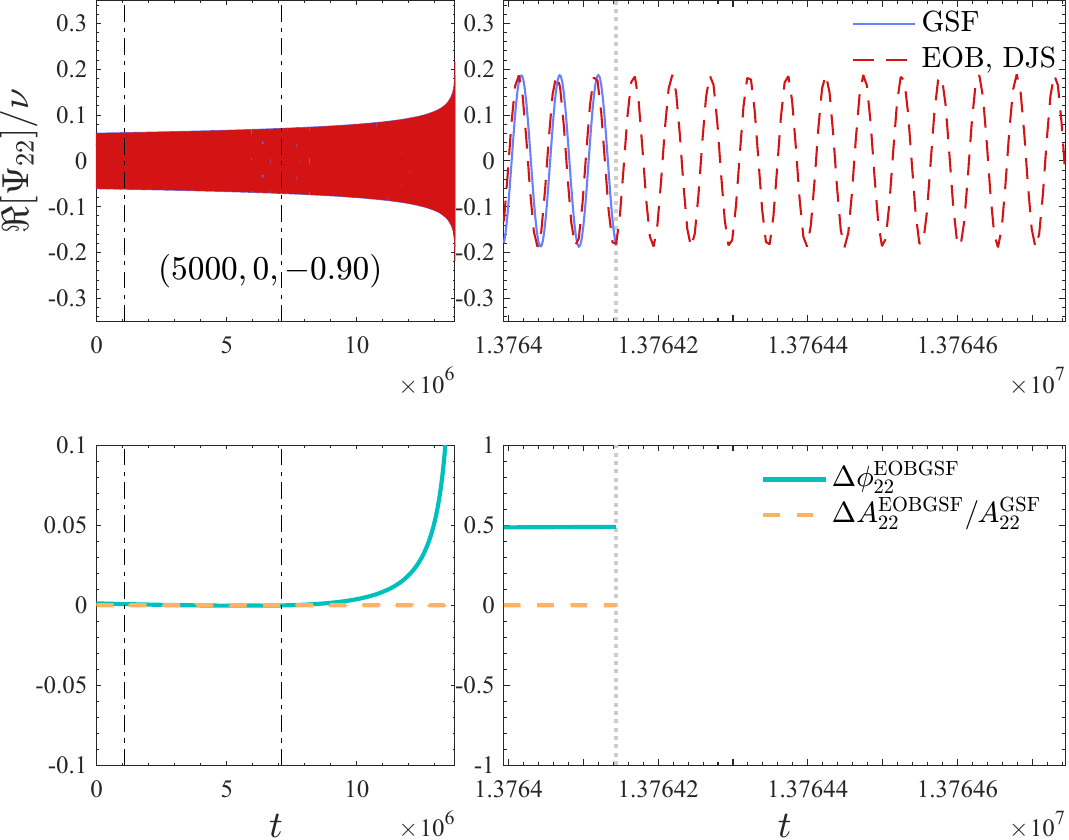}
\includegraphics[width=0.45\textwidth]{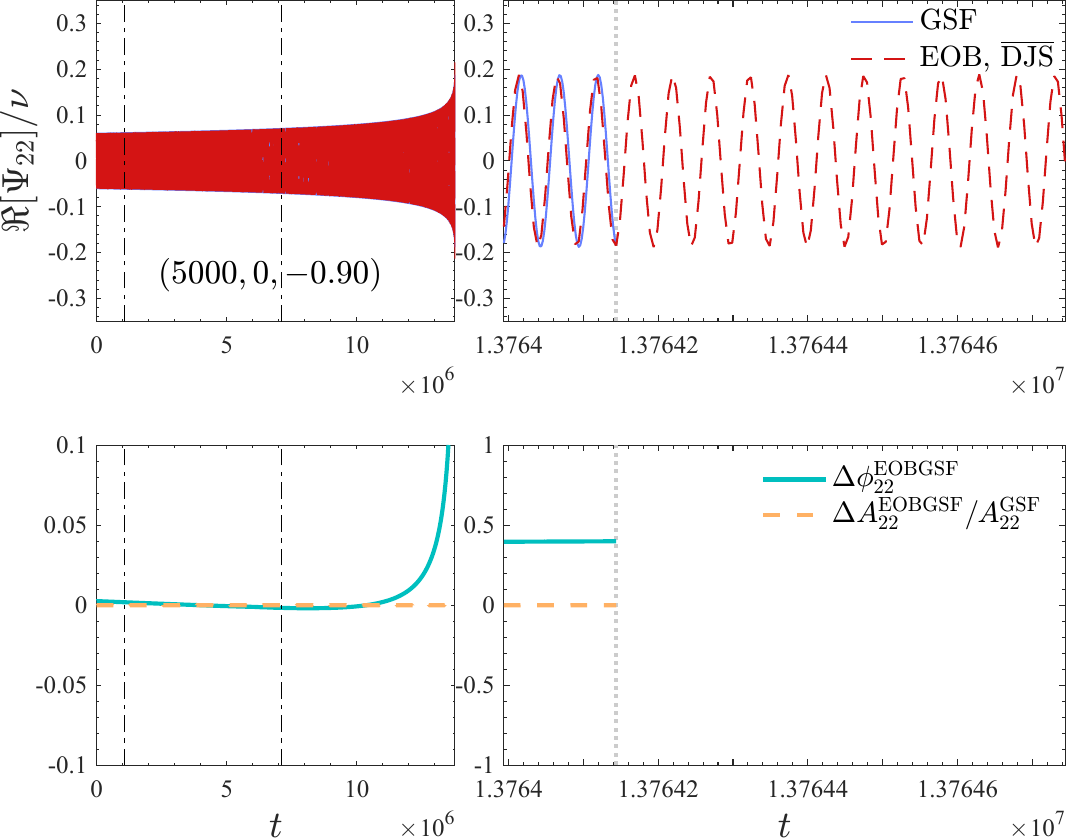}

\caption{\label{fig:phasings_aDJS} Time-domain waveform alignment for binaries with $q = 5000$, $\chi_2 = \pm 0.9$ (top/bottom), either using the standard DJS choice for
the gyro-gravitomagnetic functions (left) or their $\overline{\rm DJS}$ representation (right). The vertical dash-dotted lines in the left panels of each plot indicate the alignment interval.
In the bottom panels, the dashed orange line is the relative amplitude difference, while the turquoise line is the phase difference. Using the $\overline{\rm DJS}$ version of the gyro-gravitomagnetic functions increases the values of the dephasing for positive spins, i.e. results in the EOB waveform being even faster, while decreases it for negative spins, i.e. results in the EOB waveform being slower. Note that here the GSF waveforms are from the 1PAT1-$a$ model though the difference with the 1PAT1-$\chi$ model is very small for such a large mass-ratio.}
\end{figure*}
%


\section{Comparing the two spin gauges}
\label{sec:GSs}

The results presented in the previous section have been obtained by choosing the standard DJS gauge for the gyro-gravitomagnetic functions. In this section we consider what happens when we choose instead the $\aDJS$ gauge, with the aim of improving the agreement with GSF.

\begin{table}[htp!]
\begin{center}
\begin{ruledtabular}
\begin{tabular}{c c c c c}
$q$ & $\chi_2$ & $\omega_{\rm GSF}^{\rm end}$ & $\Delta \phi_{\rm DJS}^{\rm EOBGSF}$  &    $\Delta \phi_{\rm  \overline{DJS}}^{\rm EOBGSF}$     \\
\hline
\hline
500 &\;\;\,0.9 & $0.12329$ & $ 0.1499$  & $0.19862$ \\
	& \;\, 0.5 & $0.12406$ & $0.18843$  &  $0.20775$ \\
	&$-0.5$ & $0.12367$ & $0.34057$ & $0.32143$ \\
	&$-0.9$ & $0.1251$ & $0.42221$ & $0.37191$ \\  
5000  &\;\;\,0.9 & 0.12451& $ 0.24476$  & $0.33948$ \\
	&$-0.9$ &  0.12231 & $0.49008$ & $0.39988$ \\  
\end{tabular}
\end{ruledtabular}
\end{center} 
\caption{EOB-GSF phase differences evaluated via time-domain waveform alignment using Eq.~\ref{eq:dphi}. The third column shows the value of the GSF frequency at the end of the evolution, corresponding to the point where the phase differences are evaluated. When using the $\aDJS$ implementation of the gyro-gravitomagnetic functions, the value of the final dephasing is increasing for positive spins and decreasing for negative spins.}
\label{tab:dphi}
\end{table}

As a first test to see the impact of changing the functions $G_S, G_{S*}$, we consider waveform alignment in the time domain. In Fig.~\ref{fig:phasings_aDJS}, we show the alignment of waveforms for $q = 5000$ and $\chi_2 = \pm 0.9$, and the related phase differences are shown in the last two rows of Table~\ref{tab:dphi}. Here, the phase difference is defined as
\be
\label{eq:dphi}
\Delta \phi^{\rm EOBGSF} = \phi_{22}^{\rm EOB} - \phi_{22}^{\rm GSF}, 
\ee
and the phasing procedure is described in detail in Ref.~\cite{Albertini:2022rfe}. We see that the results yielded by changing $G_S$ and $G_{S*}$ are qualitatively different for positive and negative spins: when switching to $\aDJS$, the value of the final dephasing is increasing for positive spins and decreasing for negative spins. This is also confirmed by the other phase differences contained in Table~\ref{tab:dphi} for $q = 500$ and $\chi_2 = \pm 0.5, 0.9$.

As a side remark, we note that the alignment interval (indicated by the vertical dash-dotted lines in the left panels of each plot in Fig.~\ref{fig:phasings_aDJS}) had to be adapted very carefully for $\chi_2 = -0.9$. Aligning waveform in the time domain is an easy procedure that has been widely used in the literature to both benchmark and tune EOB waveforms with respect to NR simulations, but in some cases it can be sensitive to the alignment interval. Some of the authors recently pointed out~\cite{Nagar:2023zxh} that it is always wiser to cross-check the analytical choices in the model with some other analysis. For comparable-masses, this translates in computing the mismatch with respect to NR using different detector noises. Here, we exploit our usual $\Qo$ analysis to gain a clearer understanding of the impact of the new choice for the gyro-gravitomagnetic functions. In the following section we will see how, by considering $\Qo^{(1)}$, we can (i) understand the behaviour of the phase differences in the time-domain, and (ii) actually prove that the $\aDJS$ functions yield a closer agreement with GSF.

\section{The linear-in-spin contribution to $\Qo^{(1)}$}

\begin{figure}[t]
\includegraphics[width=0.48\textwidth]{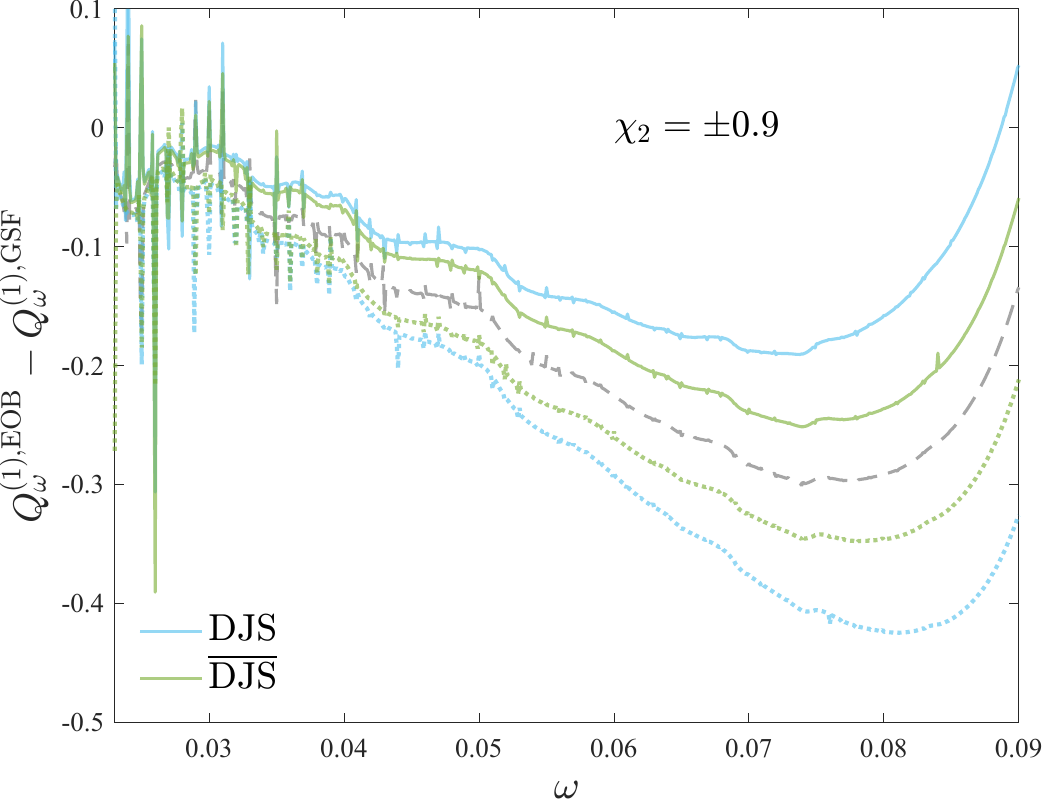} 
\caption{\label{fig:diffQ1} Difference between EOB and GSF in the 1PA term of the $\Qo$ expansion, i.e. $\Qo^{(1)}$. The dashed grey line indicates the nonspin result, while the colored lines are related to $\chi_2 = \pm 0.9$, where the dotted ones indicate the negative value of the spin. The $\aDJS$ gauge (green lines) brings the differences closer to the nonspin result, namely increases the adiabaticity for negative spins and decreases it for positive spins, allowing a better interpretation of the results obtained in the time domain.}
\end{figure}

\begin{figure}[t]
\includegraphics[width=0.48\textwidth]{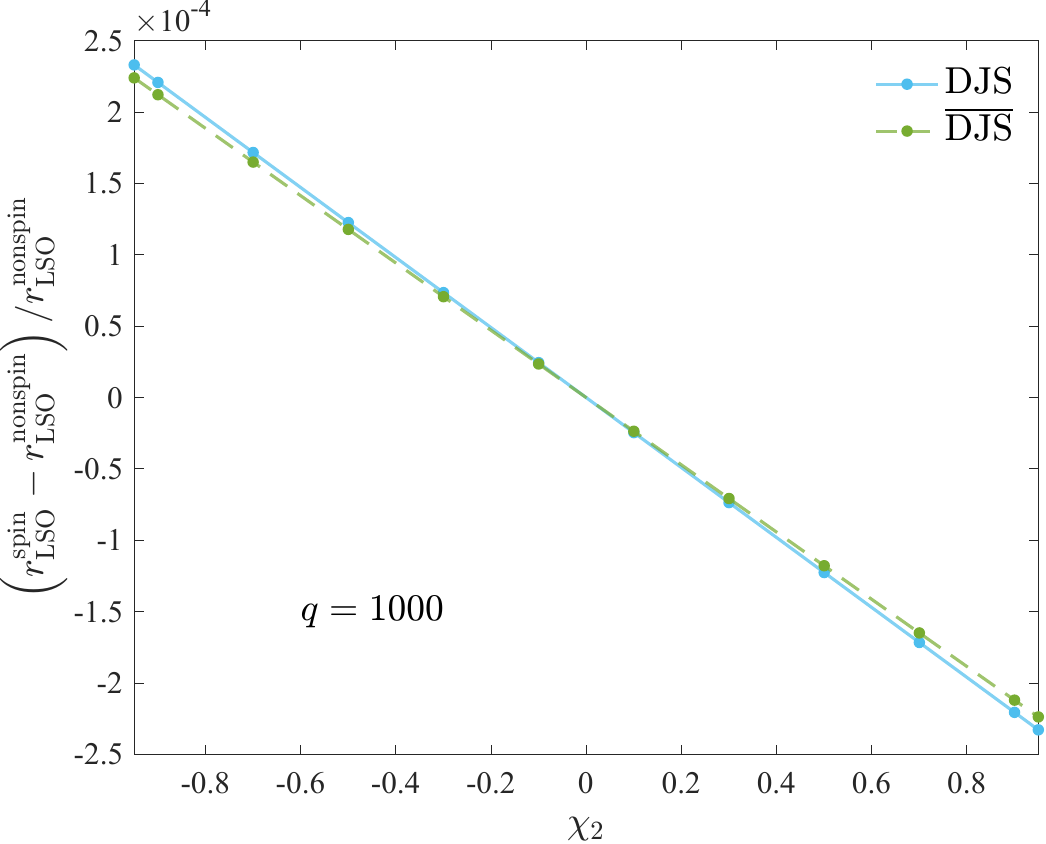} 
\caption{\label{fig:LSO} Relative difference of the LSO values for secondary spins $\chi_2 = \pm \{ 0.1, 0.3, 0.5, 0.7, 0.9, 0.95\}$ with respect to the value for nonspinning secondary obtained with (i) the gyro-gravitomagnetic functions $G_S, G_{S_*}$ in the DJS gauge, (ii) the functions in the $\aDJS$ gauge, with the complete spinning particle prefactor. The EOB LSO for a nonspinning secondary for $q = 1000$ is $r_{\rm LSO}^{\rm nonspin} = 5.9946095$. This figure shows that the second choice for the $G_S, G_{S_*}$ functions yields a larger LSO for positive spins, corresponding to a less adiabatic evolution and a faster plunge, while the opposite holds for negative spins.} 
\end{figure}

As mentioned before, the 1SF term of the $\Qo$ expansion includes a linear in the secondary spin contribution. Namely, when expanding in the symmetric mass-ratio at fixed powers of $\chi_2$ as in the 1PAT1-$\chi$ model, this term can be split as
\be
\Qo^{(1)} = \Qo^{(1,  \rm ns)} + \chi_2 \Qo^{(1, \rm s)},
\ee
where $\Qo^{(1,  \rm ns)}$ is the nonspinning part, while $\Qo^{(1, \rm s)}$ is the spinning part. Actually, the $\Qo^{(1)}$ term encapsulates all the useful spin-related information of the GSF model, since the $\Qo^{(2)}$ term is still incomplete, and  $\Qo^{(0)}$ has no dependence on the spin. In Fig.~\ref{fig:diffQ1} we plot the EOB-GSF difference in $\Qo^{(1)}$ for $\chi_2 = \pm 0.9$ using the standard gyro-gravitomagnetic functions in \TEOBResumS{} (in DJS gauge), and the new results in $\aDJS$. We see that using the $\aDJS$ functions, the EOB-GSF difference gets closer to the nonspinning result, decreasing for positive spins and increasing for negative spins. This means that EOB gets less adiabatic for positive spins and more adiabatic for negative spins. If we consider again the time-domain phase differences,
the fact that they are positive means that the EOB evolution is shorter than the GSF one at $q = 5000$. If it gets less adiabatic (even shorter), the dephasing increases, while if it gets more adiabatic (slightly longer), the dephasing decreases. This can be understood also from Fig.~\ref{fig:LSO}, that shows the last stable orbit (LSO) values obtained solving the equation $\p H_{\rm eff} / \p r = 0$ for circular orbits with the two $G_S, G_{S_*}$ versions. For positive spins, the $\aDJS$ choice corresponds to larger LSO radii and a faster plunge, while the opposite holds for negative spins.

\begin{figure}[t]
\includegraphics[width=0.48\textwidth]{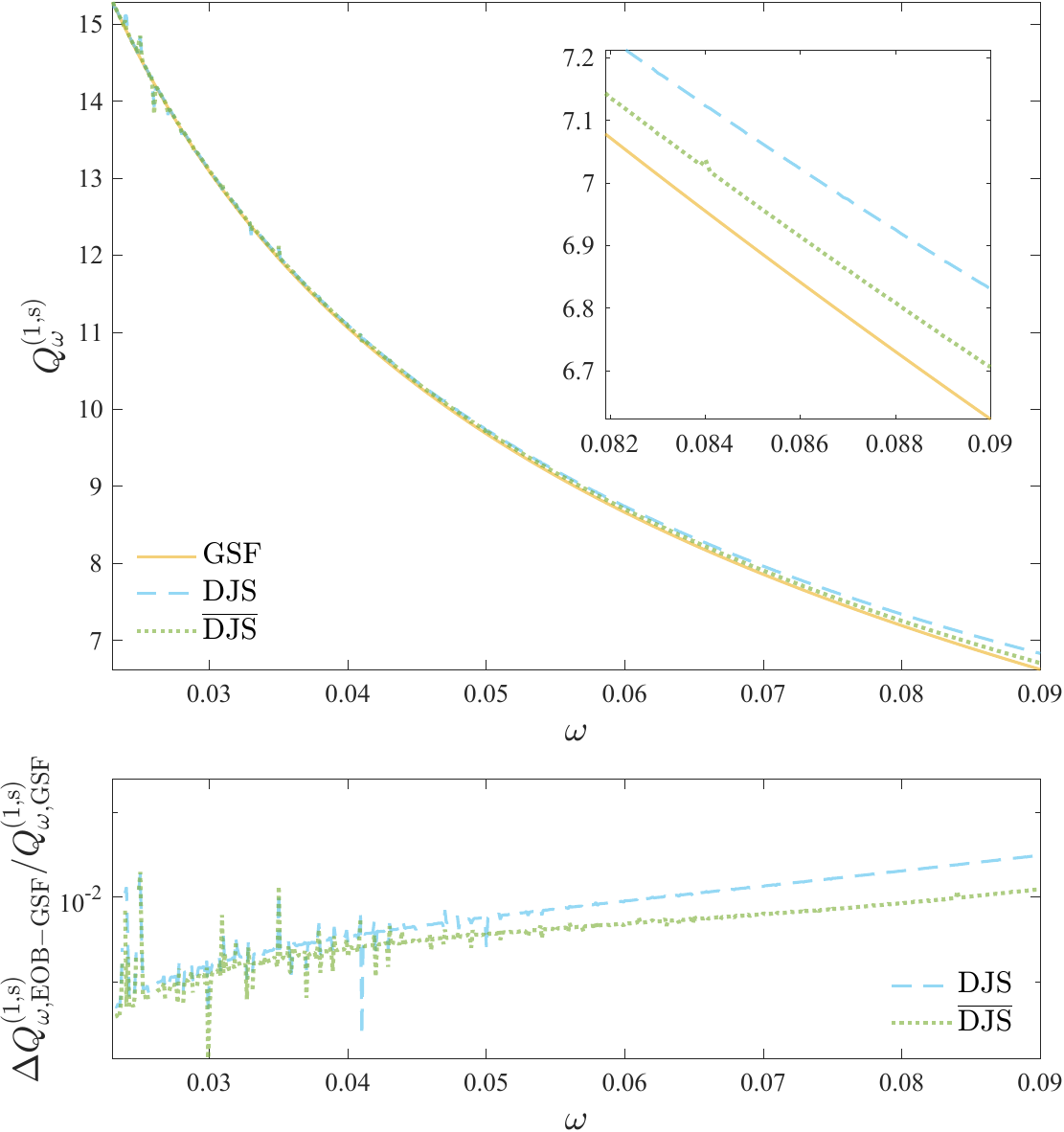} 
\caption{\label{fig:diffQ1s} Comparing the 1PA linear-in-spin contribution for EOB and GSF: plotting $\Qo^{(1, s)}$ (top) and the relative difference (bottom). For EOB, the curve is obtained as $(\Qo^{(1)} -  \Qo^{(1,  \rm ns)}) / (\chi_2) = \Qo^{(1, \rm s)}$ from the fit (using $\chi_2 = 0.9$), while the GSF contribution is extracted from the model exactly. This figure finally proves that using the $\aDJS$ gauge for the gyro-gravitomagnetic functions improves the agreement with GSF.}
\end{figure}

Hence, we see how the frequency-domain $\Qo$ analysis is helpful in interpreting the results coming from the time-domain waveform alignments. We can, however, still deepen our understanding by looking into Fig.~\ref{fig:diffQ1s}, in which we extract the spinning contribution $\Qo^{(1,s)}$ from the EOB fit and compare it to the exact GSF function. This figure shows how the EOB-GSF difference in $\Qo^{(1,s)}$ is smaller for the $\aDJS$ case, and explains in turn what we saw in Fig.~\ref{fig:diffQ1}. In fact, if we consider
\begin{align}
Q_{\omega, \rm EOB}^{(1)} - Q_{\omega, \rm GSF}^{(1)} &= Q_{\omega, \rm EOB}^{(1,\rm ns)} - Q_{\omega, \rm GSF}^{(1,\rm ns)} \nonumber \\
										       &+ \chi_2 \left( Q_{\omega, \rm EOB}^{(1,\rm s)} - Q_{\omega, \rm GSF}^{(1,\rm s)} \right) ,
\end{align}
the fact that the linear-in-spin contribution $Q_{\omega, \rm EOB}^{(1,\rm s)}$ decreases, getting nearer to the GSF result, 
indeed implies that the EOB-GSF difference in $\Qo^{(1)}$ overall gets closer to the nonspin result, i.e. explains in turn what we see in Fig.~\ref{fig:diffQ1}. 
We can integrate the dephasing due to the difference in $\Qo^{(1,s)}$, namely
\be
\label{eq:Dphi_from_Q}
\Delta\phi_{(\omega_1,\omega_2)}^{\rm EOBGSF, s} =\int_{\omega_1}^{\omega_2} \left( Q_{\omega, \rm EOB}^{(1,s)} - Q_{\omega, \rm GSF}^{(1,s)}  \right)d\log\omega \ .
\ee
For $(\omega_1,\omega_2) = (0.023,0.09)$, this calculation yields $0.086$~rad when using the DJS result, and $0.045$ when using the $\aDJS$ one. 
Only integrating the nonspinning contribution $\Qo^{(1,\rm ns)}$ on the same frequency interval yields instead $-0.204$.
Since the dephasing evaluated from Eq.~\eqref{eq:Dphi_from_Q} has to be multiplied by $\chi_2$, these results imply that for small spins, e.g. $\chi_2 = \pm 0.1$, changing the gauge has a $\sim 1\%$ impact on the 1PA contribution to the dephasing, but for larger spins, e.g. $\chi_2 = \pm 0.9$, the impact is $\sim 10\%$. We also remind here that $\Qo^{(1)}$ is the 0th-order-in-$\nu$ contribution to $\Qo$, hence these considerations are independent on the mass ratio. However, we recall that in Ref.~\cite{Albertini:2023xcn} we highlighted that for a nonspinning $q = 50000$ binary, the frequency interval $(\omega_1,\omega_2) = (0.023,0.09)$ corresponds to $\sim 13$ years if $m_2 = 10 \Msun$. By taking into account a shorter frequency interval, the integrated contribution to the phase difference at 1PA would decrease, so that the gauge choice would overall have a smaller percentage impact. 

The detailed analysis we carried on in this section showed that the new implementation of the gyro-gravitomagnetic functions in the $\aDJS$ gauge increases the agreement of the 1PA linear-in-spin contribution to the EOB conservative sector with its GSF counterpart. The impact of such choice on the phase difference is indeed more relevant for higher (absolute) values of $\chi_2$ but its relative effect also depends on the mass ratio. In particular, for more extreme mass ratios (i) the frequency interval should be chosen so that it corresponds to a meaningful length of the evolution, yielding a smaller value of the integrated dephasing, and (ii) the phase difference is mostly influenced by the 0PA term (see Eq.~\eqref{eq:Qo_nu}). Both these facts change in turn the relative influence of the 1PA linear-in-spin contribution. We still choose to keep the $\aDJS$ choice as an upgrade to our model, since we aim at producing accurate waveform templates for large mass ratios, covering both the IMRI and the EMRI regime.

%

%

\section{Conclusions}
\label{sec:conclusions}
In this work, we have carried out the first comparison of waveforms coming from the EOB and GSF approaches for black hole binaries with a spinning secondary,
and we studied how different choices for the the gyrogravitomagnetic functions $G_S$, $G_{S_*}$ entering the spin-orbit sector of \TEOBResumS{} impact the waveform.
In the standard model for comparable-mass binaries, these functions are at N2LO, NR-informed at N3LO and expressed in the DJS gauge. This gauge has been chosen following Refs.~\cite{Damour:2000we, Damour:2008qf}, and is set requiring that $G_S$ and $G_{S_*}$ do not depend on $p_\varphi$. Here we considered to use instead the $\aDJS$ gauge~\cite{Rettegno:2019tzh}, that allows to choose as prefactor in $G_{S_*}$ the EOB generalization of the full gyro-gravitomagnetic function for a spinning particle on a Kerr background. We compared the waveforms evaluated using the N3LO result in the DJS gauge of Ref.~\cite{Antonelli:2020aeb} and the recently obtained result of Ref.~\cite{Placidi:2024yld} at N3LO in the $\aDJS$ gauge. 
First, we analyzed the impact of the choice of the gauge in the time domain, by aligning the waveform and computing the accumulated phase difference. 
This procedure showed that the functions in the $\aDJS$ gauge shorten the evolution for positive spins and increase its length for negative spins. This is also confirmed by the evaluation of the LSO values, which are larger/smaller than the DJS ones for positive/negative spins.

This result alone was not clearly indicating whether the new gauge choice increased the agreement with the GSF waveforms or not, so we exploited the $\Qo$ gauge-invariant analysis to gain a better understanding. 
We fit the terms in the $\Qo$ expansion in $\nu$ for given values of the secondary spin, and then focussed on the 1PA term $\Qo^{(1)}$. When using the  $\aDJS$ gauge, the EOB-GSF difference in $\Qo^{(1)}$ decreases for positive spins and increases for negative spins. This implies that the model gets less adiabatic for positive spins and more adiabatic for negative spins, which explains the behaviour of the time-domain dephasings. By extracting the linear-in-spin contribution to $\Qo^{(1)}$ we then proved that 
the new implementation of the gyro-gravitomagnetic functions increases the agreement with GSF. In particular, the integrated dephasing on the interval $(\omega_1,\omega_2) = (0.023,0.09)$ from the EOB-GSF difference in the linear-in-spin 1PA term is roughly halved when switching from the DJS to the $\aDJS$ choice. This indeed has a greater impact for larger (absolute) values of the secondary spin and when considering mass ratios on the lower boundary of the EMRI regime. Since we aim at being as accurate as possible all over the parameter space, we choose to implement the new functions in the related branch of the public \TEOBResumSDali{} code, updating the model for large-mass-ratio inspiralling black hole binaries we presented in Ref.~\cite{Albertini:2023xcn}. 

The latter model can produce inspiral waveforms for eccentric binaries with aligned spins. Tidal effects are already included in \TEOBResumS{}, so that on this branch one can also compute inspirals of a neutron star into a massive or super-massive black hole. While on another branch of the code one can compute waveforms that are both eccentric and precessing~\cite{Gamba:2024cvy}, the interface of the latter branch with the EMRI features has not been addressed yet, and will require to be tested carefully across the parameter space. 

We conclude by stressing that, while most effects are easily incorporated into the EOB structure, it is essential to perform comparisons against other models in order to test its reliability. We are eager to keep on our program of adapting our EOB model, and to be part of a collaborative community that will be striving to get ready for the next generation of interferometers.

\acknowledgements
The authors thank Rossella Gamba for technical help with the \TEOBResumS{} public infrastructure. 
A.A. thanks Andrea Placidi and Piero Rettegno for clarifications on the results of Ref.~\cite{Placidi:2024yld} and Thibault Damour for useful discussions. J.M. thanks Adam Pound, Niels Warburton and Barry Wardell for continued helpful discussion and for sharing the second order self-force data.
A.A. and G.L.G. have been supported by the fellowship Lumina Quaeruntur No. LQ100032102 of the Czech Academy of Sciences. 
A.A. is also supported by the GAUK project No. 107324.
J.M. has been supported by the NUS Faculty of Science, under the research grant 22-5478-A0001.
The present research was also partly supported by the ``\textit{2021 Balzan Prize for Gravitation: Physical and Astrophysical Aspects}'', 
awarded to Thibault Damour. Some calculations have been performed on the Tullio server at INFN Torino.

\noindent \TEOBResumS{} is developed open source and publicly available at

{\footnotesize \url{https://bitbucket.org/eob_ihes/teobresums/src/master/}} .

The code version used in this work corresponds to the branch dev/DALI-rholm22PN and is tagged with the related arXiv number. 
The code is interfaced to state-of-art gravitational-wave data-analysis pipelines: 
\href{https://github.com/matteobreschi/bajes}{bajes}~\cite{Breschi:2021wzr}, \href{https://git.ligo.org/lscsoft/bilby}{bilby}~\cite{Ashton:2018jfp} and \href{https://pycbc.org/}{pycbc}~\cite{Biwer:2018osg}.

\appendix

\section{Analytical computation of $\Qo$ within EOB} 
\label{Qomg:anlyt}

The behavior of $\Qo$ and of its three different contributions, $\Qo^{(0)}, \Qo^{(1)}, \Qo^{(2)}$, can be addressed analytically in the circular motion limit. Assuming for simplicity that the gravitational wave frequency is $\omega_{22}=2\Omega$, where $\Omega$ is the EOB orbital frequency\footnote{We are here neglecting in the EOB waveform the additional contributions to the frequency that come from the resummed tail factor and from the residual phase correction $\delta_{22}$~\cite{Damour:2008gu}. Note that the contribution to the tail cannot be extracted analytically in closed form; see Appendix~E.2 of Ref.\cite{Nagar:2018zoe}. Nonetheless, this approximation does not change the conclusions of our reasoning provided here.}, we have
\be
\Qo =  2\frac{\Omega^2}{\dot{\Omega}} \, ,
\ee
and by using the frequency parameter  $x \equiv \Omega^{2/3}$, we have
\begin{align}
\dot{\Omega} &= \de_j \Omega \de_t j = \frac{3}{2} x^{1/2} \frac{\de x}{\de j} \hat{\mathcal{F}}_\ph \, ,
\end{align}
where $j \equiv J^{\rm circ} / \mu M$ is the orbital angular momentum along circular orbits and we employed $\de_t j =\hat{\mathcal{F}_\ph}$. The angular momentum flux is written as
$\hat{\mathcal{F}_\ph} = \mathcal{F}_\ph / \nu 
                                     =\nu \mathcal{F}_\ph^{\rm 1SF} 
                                     + \nu^2 \left( \FIISF + \chi_2 \mathcal{F}_\ph^{\rm 2SF, s} \right) 
                                     + \nu^3 \left( \FIIISF + \chi_2 \mathcal{F}_\ph^{\rm 3SF, s} \right) $
(i.e., as a 2PA expansion) to meaningfully compare EOB and GSF contributions.
Note, however, that the complete EOB flux, which is summed up to $\ell=10$, has many more 
$\nu$-dependent terms because it incorporates all the known $\nu$ dependence\footnote{The precise evaluation
of the exact $\nu$ order is tricky because of the resummed nature of the EOB fluxes. However, if we focus
only on the $\nu$ dependence of the leading, Newtonian, prefactor of each mode, $F_\lm^{\rm Newt}$, we see
that the EOB flux is at least partly $\mathcal{F}_\ph^{\rm 9SF}$.} up to 3PN order. 
The $\Qo$ function can be rewritten as
\begin{widetext}
\be
\label{Qo_exp}
\Qo(x) =\frac{4}{3} \frac{x^{5/2}}{\nu\FISF} \left\{1 - \nu \frac{ \FIISF + \chi_2 \mathcal{F}_\ph^{\rm 2SF, s} }{\FISF} 
           - \nu^2 \left[ \frac{ \FIIISF + \chi_2 \mathcal{F}_\ph^{\rm 3SF, s}  }{\FISF} - \left(\dfrac{\FIISF + \chi_2 \mathcal{F}_\ph^{\rm 2SF, s} }{\FISF}\right)^2\right]\right\} {\de_x j} .
\ee
\end{widetext}

To find the expression of the angular momentum $j(x)$ along circular orbits, we first derive its expression as a function of $u$, which is found by imposing $\p \hat{H}_{\rm eff} / \p u = 0$.
Considering that the effective Hamiltonian for circular motion is
\be
\hat{H}_{\rm eff} = A \left( 1 + p_{\varphi}^2 u_c^2  \right) + p_{\varphi} \tilde{G}, \label{eq:Horbeff} 
\ee
where we have defined $\tilde{G} \equiv G_S \hat{S} + G_{S_*} \hat{S}_*$, then $j(u)$ is given by
\begin{align}
\label{eq:ju}
j^2(u) &= - A'(u)^2 \bigg\{ -2 A(u) (\tilde{G}'(u))^2 \nonumber \\
          &+ u_c(u) A'(u) \left[ u_c(u) A'(u)+ 2 A(u) u_c'(u) \right]  \nonumber \\
          &+ 2 \sqrt{A(u)^2 \tilde{G}'(u)^2 \left[ \tilde{G}'(u)^2 - 2 u_c(u) A'(u) u_c'(u) \right]} \bigg\} ,
\end{align}
For spinning binaries, the EOB potential $A$ has the form
\be
A(u;\nu) = \frac{1 + 2u_c}{1 + 2u} \left( 1 +\nu a_{1}(u_c) \right)  \ ,
\ee
where $a_1(u_c)$ is the 1SF expression defined in Ref.~\cite{Nagar:2022fep}, evaluated as a function of the inverse centrifugal radius.
Since the 2PA expansion of the latter is $u_c = u +  \nu^2 \chi_2^2 u_c^{(2)}$, we can redefine $a_1$ and write the $A$ potential as
\be
A(u;\nu) =1- 2u +\nu a_{1}(u) + \nu^2 a_{2}(u, \chi_2) \ ,
\ee
To obtain $j(x)$, we first consider the $x(u)$ relationship 
\be
x = u + \nu U_1(u) + \nu^2 U_2(u)
\ee
that is then inverted as 
\be
\label{eq:ux}
u =x -\nu U_1(x)+\nu^2 V_2(x) \,,
\ee
where
\be
V_2(x) \equiv U_1(x) \left(\frac{d}{dx}U_1(x)\right) -U_2(x)\,.
\ee

The functions $U_1, U_2$ can be found starting from $x = \Omega^{2/3}$, where $\Omega = \p \hat{H}_{\rm EOB} / \p \varphi$ is evaluated for circular motion.
This yields
\be
x = \left[ \frac{\tG + \frac{\sqrt{A} p_\varphi u_c^2 }{\sqrt{ (1 + p_\varphi^2 u_c^2)}} }{\sqrt{1 - 2 \nu (1 - \tG p_\varphi - \sqrt{A (1 + p_\varphi^2 u_c^2)})}}\right]^{2/3} .
\ee
In this expression we substitute the angular momentum $p_\varphi$ with it circular value $j(u)$.
Finally, we consider formal expressions for the gyro-gravitomagnetic functions $G_S, G_{S_*}$
expanded up to $\nu^2$
\begin{align}
G_S &= G_S^{(0)} + \nu G_S^{(1)} + \nu^2 G_S^{(2)} , \nonumber \\
G_{S_*} &= G_{S_*}^{(0)} + \nu G_{S_*}^{(1)} + \nu^2 G_{S_*}^{(2)} ,
\end{align}
and express the spin variables $\hat{S},\hat{S}_*$ as functions of $\nu$ and $\chi_2$:
\begin{align}
\hat{S} &= \chi_2 X_2^2 = \frac{1}{4} \chi_2 (1 - \sqrt{1-4\nu})^2, \\
\hat{S}_* &= \nu \chi_2 .
\end{align}
This gives us 
\begin{widetext}
\be
U_1(u) = - \frac{1} {6 \sqrt{u} (1 - 3 u)} \bigg[ - 2 \chi_2 G_{S_*}^{(0)}(u) (1 - 3 u) ( 2 + u) + u^{3/2}\left( a_1'(u) (1-3 u) - 4 (1 - \sqrt{1-3u}) \right) + 4 u^{5/2} (3 - 2 \sqrt{1-3u}) \bigg] ,
\ee
\begin{align}
U_2(u) &= \frac{1}{144 u^2} \bigg\{ \frac{1}{1 - 3 u} \left[ u^{3/2} \left(\sqrt{1-3 u} \left(4-a_1'(u)\right)-4\right) + 2 \sqrt{1-3 u} \chi_2 \left(2 G_{S_*}^{(0)}(u) + u {G_{S_*}^{(0)}}'(u) \right)+8 u^{5/2} \right]   \nonumber \\
            & + 3 u^{3/2} \bigg[ \frac{1}{\sqrt{1-3u}} \left( 8 (1 - 2 u-\sqrt{1-3 u} )  \left(u^{3/2} a_1'(u) - 2 \chi_2 ( 2 G_{S_*}^{(0)}(u) + u {G_{S_*}^{(0)}}'(u) ) \right) \right)  \nonumber \\
            & -\frac{8 u^{3/2}}{(1-3 u)^2} \left(\sqrt{1-3 u} \left(u (2 u-1) a_1'(u) + (2-8 u) a_1(u) \right)+6 (3 u-1) \left(u \left(4 u+4 \sqrt{1-3 u}-7\right)-2 \sqrt{1-3 u}+2\right)\right)   \nonumber \\
            & + \frac{16 u \chi_2 \left(u (2 u-1)  {G_{S_*}^{(0)}}'(u) + G_{S_*}^{(0)}(u) (6 u-2)\right)}{(1-3 u)^{3/2}} + 32 \chi_2 \left( G_S^{(0)}(u) + G_{S_*}^{(1)}(u) \right) \bigg]  \bigg\}
\end{align}
\end{widetext}
where $'$ indicates derivation with respect to $u$.

By combining Eq.~\eqref{eq:ju} and Eq.~\eqref{eq:ux} we have $j(x)$, and we can finally evaluate explicitly Eq.~\eqref{Qo_exp} as a 
function of $(\FISF, \FIISF,\FIIISF,a_1(x), a_2(x))$, to obtain
\begin{widetext}
\begin{align}
\Qo^{(0)} &= - \frac{2}{3} \frac{(1- 6x)}{(1 - 3x)^{3/2}} \frac{x^{3/2}}{\FISF} \ , \\
\Qo^{(1)} &=\frac{1}{9 x^{1/2} (1-3 x)^3 {\FISF}^2} \bigg\{ \FISF \sqrt{1-3 x} \bigg[ - 2 (1 - 3 x) x \bigg( - x^{1/2} (1 - 9 x) a_1'(x)  \nonumber \\
               &+ 2 (1 - 3 x) \left(x^{3/2} a_1''(x)+\chi_2 ({G_{S_*}^{(0)}}'(x)- 2 x {G_{S_*}^{(0)}}''(x)) \right)\bigg)+3 (1-12x) x^{3/2} a_1(x)-2 \chi_2 \left(90 x^2-33 x+4\right) G_{S_*}^{(0)}(x)\bigg] \nonumber \\
               & + 2 x^{3/2} \bigg[ \FISF \left( (1-\sqrt{1-3x}-x)(1+36x^2) - 6x (1 - \sqrt{1-3x})  \right) \nonumber \\
               & - 3 \sqrt{1-3 x} (1-9x +18x^2) (\FIISF + \mathcal{F}_\ph^{\rm 2SF, s} \chi_2) \bigg] \bigg\} , \\
\Qo^{(2)} &= \frac{1}{108 x^2 (1 - 3 x)^5 {\FISF}^3} \bigg[ - 12  x^{3/2} (1 - 3 x) \FISF \cdot f\left( \chi_2, a_1(x), G_{S_*}^{(0)}(x), \FIISF, \mathcal{F}_\ph^{\rm 2SF, s}, \FIIISF, \mathcal{F}_\ph^{\rm 3SF, s} \right)   \nonumber \\
                & \quad \quad \quad \quad \quad \quad \quad \quad \quad \quad \quad \;\,  + {\FISF}^2 \cdot g \left( \chi_2, a_1(x),  a_2(x),  G_S^{(0)}(x), G_{S_*}^{(0)}(x), G_{S_*}^{(1)}(x) \right)  \nonumber \\
                & \quad \quad \quad \quad \quad \quad \quad \quad \quad \quad \quad \;\, - 72 x^3 (1-3x)^{7/2} (1 - 6x) (\FIISF + \chi_2 \mathcal{F}_\ph^{\rm 2SF, s})^2
\end{align}
\end{widetext}
where $f,g$ are long expressions that we avoid reporting here. From this computation, we can see that at 1PA the only contribution from the spin-orbit functions is $G_{S_*}^{(0)}$. This is consistent with the fact that, when expressing $\hat{S}, \hat{S}_*$ as functions of $\nu$ and $\chi_2$, the $\nu$-expansion of $\tilde{G}$ is
\be
\tilde{G} = \nu \chi_2 G_{S_*}^{(0)} + \nu^2 \chi_2 \left( G_S^{(0)} + G_{S_*}^{(1)}\right) + O(\nu^3).
\ee
This in turns justifies the meaningfulness of factorizing out the complete spinning particle function in the $\aDJS$ version of $G_{S_*}$.

\bibliography{refs20240606.bib, local.bib}

\end{document}